\documentclass[preprint,12pt]{elsarticle}
\usepackage[english]{babel}

\usepackage{amsmath, amssymb}

\usepackage{mathrsfs}

\usepackage{tensor}

\graphicspath{{figures/}}

\usepackage{color}

\newcommand{\action}{\ensuremath{\mathscr{A}}}

\newcommand{\PC}[1]{\left(#1\right)}
\newcommand{\PR}[1]{\left[#1\right]}

\newcommand{\unit}[1]{\,\text{#1}}

\journal{Physics Letters B}

\begin{document}

\begin{frontmatter}
\title{On Gravity localization under Lorentz Violation in warped scenario}
\author{Victor Santos}
\ead{victor\_santos@fisica.ufc.br}
\author{C. A. S. Almeida}
\ead{carlos@fisica.ufc.br}

\address{Physics Department, Federal University of Cear\'a,\\ %
P.O. Box 6030, 60455-760,\\ Fortaleza, Cear\'a, Brazil}

\begin{abstract}
Recently Rizzo studied the Lorentz Invariance Violation (LIV) in a brane
scenario with one extra dimension where he found a non-zero mass for the
four-dimensional graviton. This leads to the conclusion that
five-dimensional models with LIV are not phenomenologically viable. In this
work we re-examine the issue of Lorentz Invariance Violation in the context
of higher dimensional theories. We show that a six-dimensional geometry
describing a string-like defect with a bulk-dependent cosmological constant
can yield a massless 4D graviton, if we allow the cosmological constant
variation along the bulk, and thus can provides a phenomenologically viable
solution for the gauge hierarchy problem.
\end{abstract}

\end{frontmatter}

\section{Introduction}
\label{sec:introduction}

After a long time of experimental success showed by the data gathered until
now, it has been suggested that the Special theory of Relativity and the
General theory of Relativity are actually effective theories, which must be
replaced by some other theory in specific scenarios. As an example, in the
cosmological scale the energies of cosmic rays would exhibit a
Greisen-Zatsepin-Kuz'min cut off below $100\unit{EeV}$ %
\cite{1966JETPL478Z,PhysRevLett.16.748}, predicted by General Relativity,
whereas it had been detected cosmic rays above this threshold %
\cite{1995ApJ441144B}.

From the purely theoretical side, the incorporation of gravity in Quantum
Field Theory naturally leads to a minimal measurable length in the
ultraviolet regime. Some prominent approaches to Quantum Gravity, such  as
String Theory \cite{0295-5075-2-3-006,Amati198781,Amati198941} and Loop
Quantum Gravity \cite{Garay:1994en}, indicate the existence of a
minimal length of the order of the Planck length $\ell_\text{P} \sim
10^{-35}\unit{m}$, yielding a minimal length measurable. This turns to be
difficult to reconcile with the Lorentz Invariance, and then motivates
investigation of Lorentz Invariance Violation (LIV) effects.

One route to achieve LIV is by spontaneous symmetry breaking, through the
vacuum expectation value (v.e.v.) of a tensor, or in the
more common cases, a four-vector which breaks the space-time isotropy,
giving a preferred frame of reference. In such context, it was proposed by
Kostelecky and co-workers the so called Standard Model Extension (SME),
which furnishes a set of gauge-invariant LIV operators and then a framework
to investigate LIV.

Within the Standard Model (SM) there is an interesting issue, which is the
\emph{gauge hierarchy problem}. The energy scale which characterizes the
symmetry breaking for the SM is of the order of $10^{14}\unit{GeV}$ (below
of this scale the interactions are split into the electroweak and strong
ones). However, the theory of electroweak interactions (the
Glashow-Weinberg-Salam Model) itself predicts a symmetry breaking at the
scale of $10^{2}\unit{GeV}$; using the Higgs mechanism to implement such
symmetry breaking it requires a parameter tuning up to 24 digits, which
means perturbative corrections at $\mathcal{O}(10^{-24})$. This is taken as
an imperfection of the SM, and has compelled the research of the Physics
beyond it. Such a problem was addressed in the seminal work of Randall and
Sundrum (RS) \cite{Randall:1999ee}.

A framework with Lorentz Invariant Violation in an extra-dimensional
scenario had already been considered. Namely, the five-dimensional case,
with one extra \emph{flat} dimension was studied using the Standard Model
extension (SME) \cite{Appelquist:2000nn, Cheng:2002ab,Rizzo:2005um}. In
ref. \cite{Rizzo:2010vu}, Rizzo showed that LIV in a Randall-Sundrum (RS)
scenario induces a non-zero mass for the four-dimensional graviton,
resulting both from the curvature of $\text{AdS}_5$ space and the loss of
coordinate invariance. This leads to the conclusion that five-dimensional
models with LIV are not phenomenologically viable.

However, this does not forbids LIV in higher dimensional models. In this
work we shall show that a six-dimensional geometry can yield a massless 4D
graviton, if we allow the cosmological constant variation along the bulk.

\section{Randall-Sundrum gravity with LIV}
\label{sec:rs_gravity_liv}

In this section we review the work of Rizzo \cite{Rizzo:2010vu}. In
particular, we restrict ourselves to the part concerning the
Kaluza-Klein (KK) spectrum of the gravitational field.

The original setup  of the RS model is based on the Einstein-Hilbert action
\begin{equation}
  \label{eq:rs_action}
  \action = -\frac{M_5^3}{2}\int\text{d}^4x\, \text{d}y\,\sqrt{-g}\,%
  (R+2\Lambda),
\end{equation}
where $\Lambda$ is a bulk five-dimensional cosmological constant, $M_5$ is
the five-dimensional reduced Planck scale, and $y$ represents the extra
dimension coordinate. The proposed extension for the action
\eqref{eq:rs_action} which involves LIV
follows refs. \cite{PhysRevD.58.116002,PhysRevD.69.105009}:
\begin{equation}
  \label{eq:rs_action_liv}
  \action = -\frac{M_5^3}{2}\int \text{d}^4 x\,\text{d}y\, \sqrt{-g}\,%
  (R + 2\Lambda - \lambda s\indices{^{ab}}R\indices{_{ab}}),
\end{equation}
where $\lambda$ is a dimensionless constant of the order of unity which
measures the strength of the Lorentz invariance violation and
$s\indices{^{ab}} = u\indices{^a}u\indices{^b}$ is a constant tensor
defined by the v.e.v. of the 5-vector $u\indices{^a} = (0,0,0,0,1)$. The
equations of motion following from the action \eqref{eq:rs_action_liv} are
\begin{equation}
  \label{eq:eoms}
  G\indices{_{ab}} = \Lambda g\indices{_{ab}}
  +\lambda\mathcal{F}\indices{_{ab}},
\end{equation}
where $G\indices{_{ab}}$ is the Einstein tensor and
\begin{equation*}
  \mathcal{F}\indices{_{ab}} =  -\frac{1}{2} R\indices{_{cd}}\,
  s\indices{^{cd}} g\indices{_{ab}} +
  2g\indices{_{d(a}}R\indices{_{b)c}}s\indices{^{cd}}
  +\frac{1}{2}\nabla\indices{_c}\nabla\indices{_d}s\indices{^{cd}}g\indices{_{ab}}
  +\frac{1}{2}\Delta s\indices{_{ab}}
  -g\indices{_{d(a}}\nabla\indices{_{|c|}}\nabla\indices{_{|b)}}s\indices{^{cd}}
\end{equation*}
is a tensor arising from the LIV term in the \eqref{eq:rs_action_liv}. The
parenthesis bracketing the indices denotes symmetrization,
\[
  T\indices{_{(ab)}} = \frac{1}{2}\PC{T\indices{_{ab}}
  + T\indices{_{ba}}}
\]
and $\Delta$ denotes the Laplace operator
$\Delta=\nabla\indices{^a}\nabla\indices{_a}$.

In order to investigate the implication of such additional term in the
four-dimensional effective field theory, we need first identify the massless
gravitational fluctuations around the background solution. Since in this case
the background solution is
\begin{equation}
  \label{eq:background_solution_rs_model}
  \text{d}s^2 = e^{-2ky}\eta\indices{_{\mu\nu}}\text{d}x^\mu\text{d}x^\nu
  -\text{d}y^2,
\end{equation}
we take tensor fluctuations of the form
\begin{equation}
  \label{eq:tensor_fluctuations}
  \text{d}s^2 = e^{-2ky}(\eta\indices{_{\mu\nu}}
  +\gamma\indices{_{\mu\nu}})\text{d}x^\mu\text{d}x^\nu
  -\text{d}y^2,
\end{equation}
where $\gamma\indices{_{\mu\nu}}$ represents the physical graviton in
the four-dimensional theory. Therefore, the massless gravitational
fluctuations are identified by the resulting linearized equations obtained
from eq.~\eqref{eq:eoms}.

In the case of a \emph{flat} five-dimensional background, the derivation
was performed by Carroll and Tam \cite{Carroll:2008pk}, and it was found
the equations of motion
\begin{equation}
  \label{eq:eom_flat_case}
  \Box\gamma\indices{_{\mu\nu}} = \lambda\,\partial_y^2\gamma\indices{_{\mu\nu}}.
\end{equation}
This gives the Kaluza-Klein masses $m_n^2 = n^2(1+\lambda)/R^2$, and
wavefunctions of trigonometric form. This result agrees with the results
found for scalar and fermionic fields, already discussed in ref.
\cite{Rizzo:2010vu}. It is worthwhile to notice that the LIV parameter
$\lambda$ does not leads to a bulk mass term in the equation of motion,
and thus it does not have any inconsistency with the 4D behavior for the
graviton.

In curved spacetime however, the picture changes. The equation of motion
for the $yy$ component is non-dynamical and relates the cosmological
constant to the LIV parameter $\lambda$:
\begin{equation*}
  \Lambda = -6 k^{2} \PC{1 + \frac{\lambda}{3}}.
\end{equation*}
The usual RS result is clearly obtained in the limit $\lambda\rightarrow
0$. The difference is that since the value of $\lambda$ varies, its decreasing
to large negative values causes a topology change
$\text{AdS}_5\rightarrow\text{dS}_5$. Then, for consistency with the RS
model there must be imposed the condition $\lambda > -3$. Moreover, in this
case there are no brane tensions to tune, and thus the other equations of
motion are automatically satisfied.

The linearized equations of motion for the tensor fluctuations
$\gamma\indices{_{\mu\nu}}$ are given by
\begin{equation}
  \label{eq:linearized_eoms_rs}
  -\Box\gamma\indices{_{\mu\nu}}
  + e^{-2ky}(1+\lambda)\partial_y^2\gamma\indices{_{\mu\nu}}
  - 4ke^{-2ky}(1+\lambda)\partial_y\gamma\indices{_{\mu\nu}}
  - 8k^2 \lambda e^{-2ky}\gamma\indices{_{\mu\nu}} = 0
\end{equation}
and under the Kaluza-Klein decomposition
\begin{equation}
  \label{eq:kk_decomposition}
  \gamma\indices{_{\mu\nu}}(x^\rho,y) = \sum_{n=0}^{\infty}M^{(n)}_{\mu\nu}(x^\rho)\phi_{n}(y),
\end{equation}
where we require $\Box M^{(n)}_{\mu\nu}=-m_n^2 M^{(n)}_{\mu\nu}$, we can
find the graviton equations of motion, namely
\begin{equation}
  \label{eq:eom_curved_space}
  -\partial_y\PC{e^{-4ky}\,\partial_y\phi_n} + m^{2}e^{-4ky}\,\phi_n
  = \tilde{m}_{n}^{2}e^{-2ky}\,\phi_n,
\end{equation}
where $\tilde{m}_n^2 = m_n^2/(1+\lambda)$ and
\begin{equation}
  \label{eq:bulk_mass}
  m^2 = 8k^2\frac{\lambda}{1+\lambda}.
\end{equation}
From eq.~\eqref{eq:eom_curved_space} we can see that, unlikely the flat
case, the curvature of the $\text{AdS}_5$ space (encoded in the
constant $k$) together with the Lorentz violation
led to a bulk mass term \eqref{eq:bulk_mass},
\begin{equation*}
  m^2 = 8k^2\frac{\lambda}{1+\lambda},
\end{equation*}
which is zero only when $\lambda=0$, that is, when there is no Lorentz
violation. Therefore the presence of LIV prevents the existence of a
massless KK zero mode, leading to a phenomenological inconsistency
since the zero mode is associated to the ordinary graviton in 4D. Since
the bulk graviton mass depends on the nature of the background metric, in
the next section we shall construct an example of a curved metric which
yields a massless KK zero-mode.

\section{Gravity on a string-like defect with LIV}
\label{sec:gravity_stringlike_defect}

Now we shall see how gravity behaves in a brane with specific structure. As
our geometric background, we choose a local string defect, which was
already studied in ref.~\cite{Gherghetta:2000qi}.

\subsection{String-like defect: Einstein's equations and consistency relations}

Let us recall the geometric setup of~\cite{Gherghetta:2000qi}. In six
dimensions, we start from the field equations
\begin{equation}
  \label{eq:eoms_6d_case}
  G\indices{_{ab}} = \Lambda g\indices{_{ab}}
  + \frac{1}{M_6^4} T\indices{_{ab}},
\end{equation}
where $\Lambda$ is the bulk cosmological constant
$T\indices{_{ab}}$ is the stress-energy tensor
and $M_6$ is the bulk reduced mass scale.
Also, we assume the existence of a six-dimensional geometry which respects
the Poincar\'e invariance in the 4D brane, namely
\begin{equation}
  \label{eq:metric_ansatz}
  \text{d}s^{2} = A(\rho)\eta\indices{_{\mu\nu}}
  \text{d}x\indices{^\mu}\text{d}x\indices{^\nu}
  - \text{d}\rho^2
  - R_0^2\,B(\rho)\text{d}\theta^2,
\end{equation}
where the two extra dimensions are chosen to be polar coordinates
$\rho\in[0,\infty)$ and $\theta\in[0,2\pi)$.

The metric ansatz \eqref{eq:metric_ansatz} yields the following
general expression for the 4D mass scale $M_\text{P}$:
\begin{equation}
  M_\text{P}^2 = 2\pi M_6^4\int_0^\infty\text{d}\rho\, A\sqrt{B}.
\end{equation}
The nonzero components for the stress-energy tensor are assumed to be
\begin{equation}
  \label{eq:stress_tensor}
  T\indices{^\mu_\nu} = \delta\indices{^\mu_\nu}\tau_0,\quad
  T\indices{^r_r} = \tau_r,\quad
  T\indices{^\theta_\theta} = \tau_\theta,
\end{equation}
where the source functions $\tau_0$, $\tau_r$ and $\tau_\theta$ are all
radial dependent. Then, using \eqref{eq:metric_ansatz} and
\eqref{eq:stress_tensor} the equations of motion are
\begin{align}
  \frac{3}{2}\frac{A^{\prime\prime}}{A}
  + \frac{3}{4}\frac{A^\prime}{A}\frac{B^\prime}{B}
  - \frac{1}{4}\PC{\frac{B^\prime}{B}}^2
  + \frac{1}{2}\frac{B^{\prime\prime}}{B} &=
  -\PC{\Lambda + \frac{\tau_0}{M_6^4}},\label{eq123}\\
  \frac{3}{2}\PC{\frac{A^\prime}{A}}^2
  + \frac{A^\prime}{A}\frac{B^\prime}{B}
  &= -\PC{\Lambda + \frac{\tau_r}{M_6^4}},\label{456}\\
  2\frac{A^{\prime\prime}}{A}
  + \frac{1}{2}\PC{\frac{A^\prime}{A}}^2 &=
  -\PC{\Lambda + \frac{\tau_\theta}{M_6^4}},\label{eq:eomtheta}
\end{align}
where the prime ($\prime$) denotes differentiation with respect to $\rho$.
The sources are related to the geometry through the continuity equation
\begin{equation}
\label{eq:continuity_equation}
\tau^\prime_r
-2\frac{A^\prime}{A}\PC{\tau_0-\tau_r}
+\frac{1}{2}\frac{B^\prime}{B}\PC{\tau_r-\tau_\theta} = 0.
\end{equation}
One can think now that there is a local string-like 3-brane at
$\rho=0$, having a nonvanishing radial stress-energy
\eqref{eq:stress_tensor}, like the Nielsen-Olesen string solution in the
6D Abelian Higgs model. The source functions describe then a matter
distribution which is zero except within a core of radius $\epsilon$, and
we can define the components of the string tension per unit length as
\begin{equation}
  \mu_i = \int_0^\epsilon \text{d}\rho\, A^2\sqrt{B}\,\tau_i(\rho),
\end{equation}
where $i=0,\rho,\theta$.

\subsection{LIV in the 6D string-like bulk}
In order to investigate LIV in a six-dimensional scenario, we will follow a
``conservative'' (i.e., the same) approach used in ref. \cite{Rizzo:2005um} and choose a
similar action to \eqref{eq:rs_action_liv}:
\begin{equation}
  \label{eq:stringlike_action_liv}
  \action = -\frac{M_6^2}{2}\int\text{d}^4x\text{d}r\text{d}\theta\,
  \sqrt{-g}\PC{R + 2\Lambda - \lambda s\indices{^{ab}}R\indices{_{ab}}},
\end{equation}
where $u^{A}=(0,0,0,0,1,0)$ and $\lambda$ is as discussed in section
\ref{sec:rs_gravity_liv}. This action acts like in the 5D case, where the
Lorentz invariance is broken in the extra dimension (only the $r$ direction
in this case), whereas it is preserved in 4D. Although we could have in
principle a broader set of LIV operators (for example, one could consider
$u^A=(0,0,0,0,1,1)$), the physical implications of such terms go beyond
the scope of this text, and they will be considered in detail in the study
of the KK spectrum, to be discussed in a future work.

Action \eqref{eq:rs_action_liv} gives the equations of motion
\begin{align}
  &\frac{3}{2}\frac{A^{\prime\prime}}{A}
  + \frac{3}{4}\frac{A^\prime}{A}\frac{B^\prime}{B}
  - \frac{1}{4}\PC{\frac{B^\prime}{B}}^2
  + \frac{1}{2}\frac{B^{\prime\prime}}{B} =
  -\frac{1}{1+\lambda}\PC{\Lambda + \frac{\tau_0}{M_6^4}},\\
  &-2\lambda\frac{A^{\prime\prime}}{A}
  + (1+\lambda)\frac{A^\prime}{A}\frac{B^\prime}{B}
  +\PC{\frac{3}{2} + \frac{5\lambda}{2}}\PC{\frac{A^\prime}{A}}^2
  + \frac{\lambda}{4}\PC{\frac{B^\prime}{B}}^2
  - \frac{\lambda}{2}\frac{B^{\prime\prime}}{B} =
  -\PC{\Lambda + \frac{\tau_r}{M_6^4}},\\
  &2\frac{A^{\prime\prime}}{A}
  + \frac{1}{2}\PC{\frac{A^\prime}{A}}^2 =
  -\frac{1}{1+\lambda}\PC{\Lambda + \frac{\tau_\theta}{M_6^4}},
\end{align}
and the continuity equation
\begin{equation}
6\tau^\prime_r
-2\frac{A^\prime}{A}\PC{\tau_0-\tau_r}
+\frac{1}{2}\frac{B^\prime}{B}\PC{\tau_r-\tau_\theta} =
-M_6^4(\Lambda^\prime+\lambda\mathcal{T}),
\label{eq:continuity_equation}
\end{equation}
where
\begin{align*}
\mathcal{T}(\rho) &= -25\PC{\frac{A^\prime}{A}}^3
-\frac{9}{2}\PC{\frac{A^\prime}{A}}^2\frac{B^\prime}{B}
+\frac{11}{4}\PC{\frac{B^\prime}{B}}^2
-\frac{23}{8}\PC{\frac{B^\prime}{B}}^3\nonumber\\
&\phantom{=\ }+35\frac{A^\prime}{A}\frac{A^{\prime\prime}}{A}
+4\frac{A^{\prime\prime}}{A}\frac{B^\prime}{B}
-3\frac{B^{\prime\prime}}{B}\PC{1-\frac{B^\prime}{B}}\nonumber\\
&\phantom{=\ }-\frac{A^\prime}{A}\PR{\frac{B^\prime}{B}
 + \frac{9}{2}\PC{\frac{B^\prime}{B}}^2 - 5\frac{B^{\prime\prime}}{B}}
-12\frac{A^{\prime\prime\prime}}{A}.
\end{align*}

Note that we allow a radial dependence $\Lambda=\Lambda(\rho)$, which can
be interpreted as a fluid with variable density in the  bulk, due to a
matter distribution within the core of string.

Now, in order to simplify our construction we will seek for a
solution where $A(\rho) = B(\rho)$. In this case, after a tedious
algebra the linearized equations are
\begin{eqnarray}
  \label{eq:linearized_equations_string}
  (1+\lambda)\partial_r^2\gamma
  + \frac{(5+7\lambda)}{2} \frac{A^\prime}{A} \partial_r\gamma
  -\PR{2\Lambda + \PC{\frac{3}{2}+5\lambda}\PC{\frac{A^\prime}{A}}^2 + 4\PC{1 + \frac{3\lambda}{2}}\frac{A^{\prime\prime}}{A}}\gamma \nonumber \\= \frac{1}{A}\Box\gamma - \frac{1}{R_0^2 A}\partial_\theta^2\gamma.
\end{eqnarray}
For sake of simplifying our investigation, we tune $\tau_\theta = 0$, which is
equivalent to take the string tension
\begin{equation*}
  \mu_\theta = \int_0^\epsilon \text{d}\rho\,A^{3/2}\,\tau_\theta(\rho)
\end{equation*}
to zero. In this way, choosing $\tau_\theta = 0$, equation
\eqref{eq:eomtheta} is simply a definition of the cosmological ``constant'' in terms
of the metric factor
\begin{equation}
  \label{eq:cosmological_constant}
  -\frac{\Lambda}{1+\lambda}  =  2\frac{A^{\prime\prime}}{A}+ \frac{1}{2}\left(\frac{A^\prime}{A}\right)^2.
\end{equation}
Therefore, under the usual Kaluza-Klein decomposition
$\gamma\indices{_{\mu\nu}}(x\indices{^\lambda},r,\theta) =
M\indices{_{\mu\nu}}(x\indices{^\lambda})\psi(\rho)e^{\text{i}\ell\theta}$,
where we require as usual that $(\Box + m^2_0)M\indices{_{\mu\nu}} = 0$, we
get from eq. \eqref{eq:linearized_equations_string} the graviton
wavefunction equations
\begin{equation}
  \psi_m^{\prime\prime}
  + \frac{(5+7\lambda)}{2(1+\lambda)} \frac{A^\prime}{A} \psi^{\prime}_m
  -\PR{\frac{4+11\lambda}{2+2\lambda}\PC{\frac{A^\prime}{A}}^2
  -\frac{4 + 8\lambda}{2+2\lambda}\frac{A^{\prime\prime}}{A}}\psi_m = -\frac{1}{A}\frac{m^2}{1+\lambda}\psi_m,
\end{equation}
where $m^2 = m_0^2 - l^2/R_0^2$.

From this last equation we can see that a zero bulk mass can be achieved only if
\begin{equation}
  \label{eq:bulk_mass_equation}
  \frac{A^{\prime\prime}}{A} - \frac{4 + 11\lambda}{4+8\lambda}\PC{\frac{A^\prime}{A}}^2 = 0.
\end{equation}
Note that in the case where we have $\lambda = 0$ this equality is
automatically satisfied by the exponential solution $A(\rho) = e^{-k\rho}$,
where
\begin{equation}
  k = \sqrt{\frac{2}{5}\frac{(-\Lambda)}{M_6^4}}
\end{equation}
already found in ref. \cite{Gherghetta:2000qi}. Also, in this case
$\Lambda$ is a simple numerical constant.

In terms of the warp factor $w(\rho)$ defined by  $A(\rho) = e^{w(\rho)}$,
equation \eqref{eq:bulk_mass_equation} becomes
\begin{equation*}
  w'' - \tilde{\lambda}(w^\prime)^2 = 0,\quad \tilde{\lambda}=\frac{3\lambda}{4+8\lambda}.
\end{equation*}
This is a Riccati equation for $w^\prime$, and its solution is
\begin{equation}
  \label{eq:warp_factor}
  w(\rho) = w_0 - \frac{1}{\tilde{\lambda}}\log\PC{1 +
  c\rho\tilde{\lambda}},
\end{equation}
where $w_0$ and $c$ are integration constants, to be chosen by the
boundary conditions imposed on the metric coefficients.
From this we can see that a consistent solution for any $\rho$ can be found if we impose
$\lambda > 0$.

\begin{figure}[!htb]
  \centering
  \includegraphics[scale=0.5]{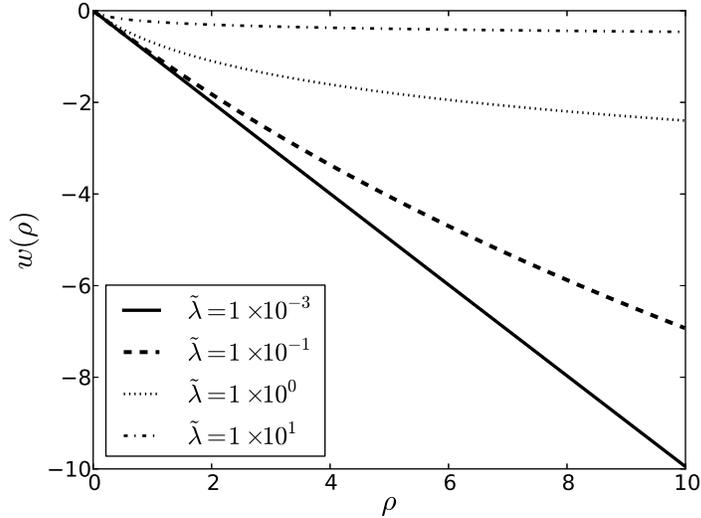}
  \caption{Warp factor for the string-like geometry for some values of $\tilde{\lambda}$ and $c=1$. Notice the usual result (linear behavior) being recovered when the LIV parameter approaches zero.}
  \label{fig:warp_factor}
\end{figure}

Despite the odd appearance of this warp factor, a power series expansion near the origin $\rho=0$, namely
\begin{equation*}
  w(\rho) = w_0 - c\rho + \frac{c^2\tilde{\lambda}}{2}\rho^2 + \frac{c^3 \tilde{\lambda}^2}{3}\rho^3,
\end{equation*}
reveals that we recover the usual result for a string-like defect either
when $\lambda\rightarrow 0$ or when $\rho\rightarrow 0$, when we
identify $c\rightarrow k$. The behavior of
the expression \eqref{eq:warp_factor} can be observed in figure
(\ref{fig:warp_factor}).

\begin{figure}[!htb]
  \centering
  \includegraphics[scale=0.5]{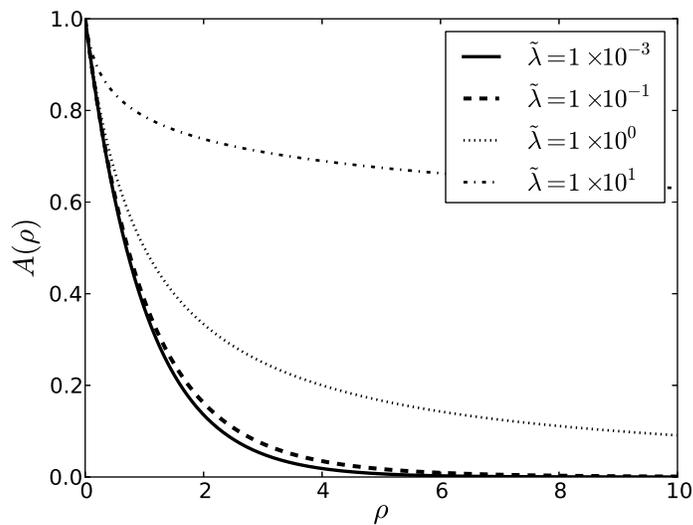}
  \caption{Metric function for the string-like geometry for some values of $\tilde{\lambda}$ and $c=1$.}
  \label{fig:exp_warp_factor}
\end{figure}

The metric function is given by
\begin{equation}
  \label{eq:metric_function}
  A(\rho) = A_0(1 + c\rho\,\tilde{\lambda})^{-1/\tilde{\lambda}},
\end{equation}
where $A_0 = A(0)$. Its behavior is shown in figure (\ref{fig:exp_warp_factor}).

The cosmological constant in our case is given by
\begin{equation}
  -\Lambda(\rho) = \frac{8c^2(26\lambda^3+49\lambda^2+28\lambda+5)}%
  {16[4\lambda(\lambda+1)+1]+[3(3c+16)\lambda^2+24\lambda]c\rho},
\end{equation}
which reduces to a constant $(-\Lambda) = 5c^2/2$ in the Lorentz Invariant case, as we would expect.

Here it is worth to mentioning that an argument that favors the string-like geometry in contrast to 5D models concerning the massless graviton lies in the bulk curvature. Indeed, in the Rizzo work \cite{Rizzo:2010vu} the curvature is constant (AdS space of the Randall-Sundrum model). Nevertheless, we can straightforwardly to get a following expression for the bulk curvature in our case (string-like geometry), namely
\begin{equation}
  \label{eq:6d-curvature}
R(\rho) = \frac{5c^2}{2}\frac{3+2\tilde{\lambda}}{(1+c\rho\tilde{\lambda})^2}.
\end{equation}

Note that this curvature depends on the radial coordinate. On the other hand, we recall the eq. \eqref{eq:bulk_mass}, where the curvature of
the $\text{AdS}_5$ space (encoded in the constant $k$) together with the Lorentz violation led to a bulk mass term different from zero. It is reasonable to assume that in our case this non-constant curvature can be joining with the Lorentz violation contribution in order to annul the graviton mass.

\section{Four-dimensional effective mass scale}
\label{sec:brane_mass_scale}
The geometry of string-like models is richer than the 5D RS type 2 model,
since the exterior space-time of the string-brane is conical with deficit angle proportional to
the string tensions \cite{Gherghetta:2000qi,Olasagasti:2000gx,Chen:2000at,Giovannini:2001hh}. Therefore, the exterior geometry of the string-brane expresses
the physical content of the brane. Now we wish to study how the Lorentz invariant violation and the new 6D geometry alters the
relation between the bulk and brane mass scale.

Note that all the metric components are limited functions, hence this geometry has a finite volume
and then, it can be used to tuning the ratio between the Planck masses explaining the
hierarchy between them. Therefore, in order to obtain the relationship between the four-dimensional Planck mass (M4) and
the bulk Planck mass (M6) let us now consider the action \eqref{eq:stringlike_action_liv}, under the
stringlike solution characterized by the metric factor \eqref{eq:metric_function}, namely,
\begin{equation}
  \action = -\frac{M_6^4}{2}\int\text{d}^4x\text{d}r\text{d}\theta\,
  \sqrt{-g}\PC{R + 2\Lambda - \lambda R\indices{_{44}}}.
\end{equation}

Comparing with the effective four-dimensional action
\begin{equation}
  \action_\text{brane}=-\frac{M_\text{P}^2}{2}\int_\text{brane}
  \text{vol}_\text{brane}(R_4+2\Lambda_4),
\end{equation}
we can find the relation between the mass scale in the bulk and the brane as
\begin{equation*}
  \frac{M_\text{P}^2}{M_6^4}=2\pi R_0\int_0^\infty\text{d}\rho\,
  A^{3/2}.
\end{equation*}
Since the metric factor is given by \eqref{eq:metric_function} we find
\begin{equation*}
  \frac{M_\text{P}^2}{M_6^4}=2\pi R_0
  \left[\frac{2 A_0^{3/2}}{c(2\tilde{\lambda}-3)}(1+c\rho\tilde{\lambda})^%
  {(2\tilde{\lambda}-3)/2\tilde{\lambda}}\right]_0^\infty
  = \frac{4\pi R_0 A_0^{3/2}}{c(3-2\tilde{\lambda})},
\end{equation*}
or
\begin{equation}
  M_\text{P}^2=
  \frac{8\pi R_0 A_0^{3/2}}{3c}
  \left(\frac{2\lambda+1}{3\lambda+2}\right)M_6^4.
\end{equation}

\section{Summary and conclusions}

We constructed a warped six-dimensional geometry which realized Lorentz
Invariance Violation (LIV) with a usual four-dimensional graviton. Although
the construction has the side effect of yield a radial dependent
cosmological constant, the model provides a framework to study Lorentz
Invariance violation in warped scenarios on physical grounds. Indeed, we
obtain a massless four-dimensional graviton. Some directions to be taken
are: (i) study another LIV terms besides the one involving the Ricci tensor
(like
$u\indices{^a}u\indices{^b}u\indices{^c}u\indices{^d}R\indices{_{abcd}}$),
and investigate the conditions for a massless graviton; (ii) study the mass
spectra in this framework, enabling one to find its contribution to the
gravitational potential and possibly find experimental upper bounds for the
LIV parameter $\lambda$. Also, another question is how our solution can be
found in an effective supergravity theory.

\section{Acknowledgments}

The authors thank the financial support of CNPq and CAPES (Brazilian Agencies).

\section*{References}
\bibliographystyle{model1-num-names}
\bibliography{liv}







\end{document}